%% file: Catalog_oEA6.tex
\documentclass[a4paper,numbers]{raa}            
\usepackage{lscape,longtable}
\usepackage[numbers]{natbib}            
\usepackage[dvips,colorlinks=true, linkcolor=red,urlcolor=magenta,citecolor=blue,anchorcolor=green]{hyperref}
\usepackage{graphicx,times}             
\usepackage{deluxetable}
\usepackage{color}
\def\vp{\color{blue}}
\definecolor{MyDarkBlue}{rgb}{0.1,0,0.55}

\bibpunct[,]{[}{]}{,}{n}{}{,}      

\begin{document}

   \title{Pulsating Components in Binary and Multiple Stellar Systems ---
   A Catalog of Oscillating Binaries
\,$^*$
\footnotetext{$*$ An update to arXiv:1002.2729v5 based on the literature as of
October 26, 2014. The web version at~\url{http://www.chjaa.org/COB/} provides the latest updates.
This edition is dedicated to my wife JYZ who suggests me to make the catalog searchable, sortable,
and easy to update.
}}

   \volnopage{Vol.15 (2015) No.?, 000--000}      
   \setcounter{page}{1}          

   \author{A.-Y. Zhou
   }

   \institute{National Astronomical Observatories, Chinese Academy of Sciences,
             Beijing 100012, China; {\it aiying@nao.cas.cn}\\
   }

   \date{Received~~2010 month day; accepted~~2010~~month day}

\abstract{ We present an up-to-date catalog of pulsating binaries,
i.e. the binary and multiple stellar systems
containing pulsating components, along with a statistics on them.
Compared to the earlier compilation by Soydugan et al.(2006a) of
25 $\delta$ Scuti-type `oscillating Algol-type eclipsing binaries' (oEA),
the recent collection of 74 oEA by Liakos et al.(2012),
and the collection of Cepheids in binaries by Szabados (2003a),
the numbers and types of pulsating variables in binaries are now extended.
The total numbers of pulsating binary/multiple stellar systems have increased to be 515
as of 2014 October 26, among which 262+ are oscillating eclipsing binaries and
the oEA containing $\delta$ Scuti components are updated to be 96.
The catalog is intended to be a collection of various pulsating binary stars
across the Hertzsprung-Russell diagram.
We reviewed the open questions, advances and prospects connecting pulsation/oscillation and binarity.
The observational implication of binary systems with pulsating components,
to stellar evolution theories is also addressed.
In addition, we have searched the Simbad database for candidate pulsating binaries.
As a result, 322 candidates were extracted. Furthermore, a brief statistics on
Algol-type eclipsing binaries (EA) based on the existing catalogs is given.
We got 5315 EA, of which there are 904 EA with spectral types A and F.
The present catalog has a sortable web version allowing easy updating and maintenance (\url{http://www.chjaa.org/COB/}).
}

   \authorrunning{A.-Y. Zhou}            
   \titlerunning{Catalog of Oscillating Binaries (\url{http://www.chjaa.org/COB/})}  

   \maketitle

%
%
\section{Motivation}           
\label{sect:intro}

\begin{figure}[t!!!]
   \vspace{2mm}\centering
   \includegraphics[width=140mm, height=90mm,angle=0,scale=0.99]{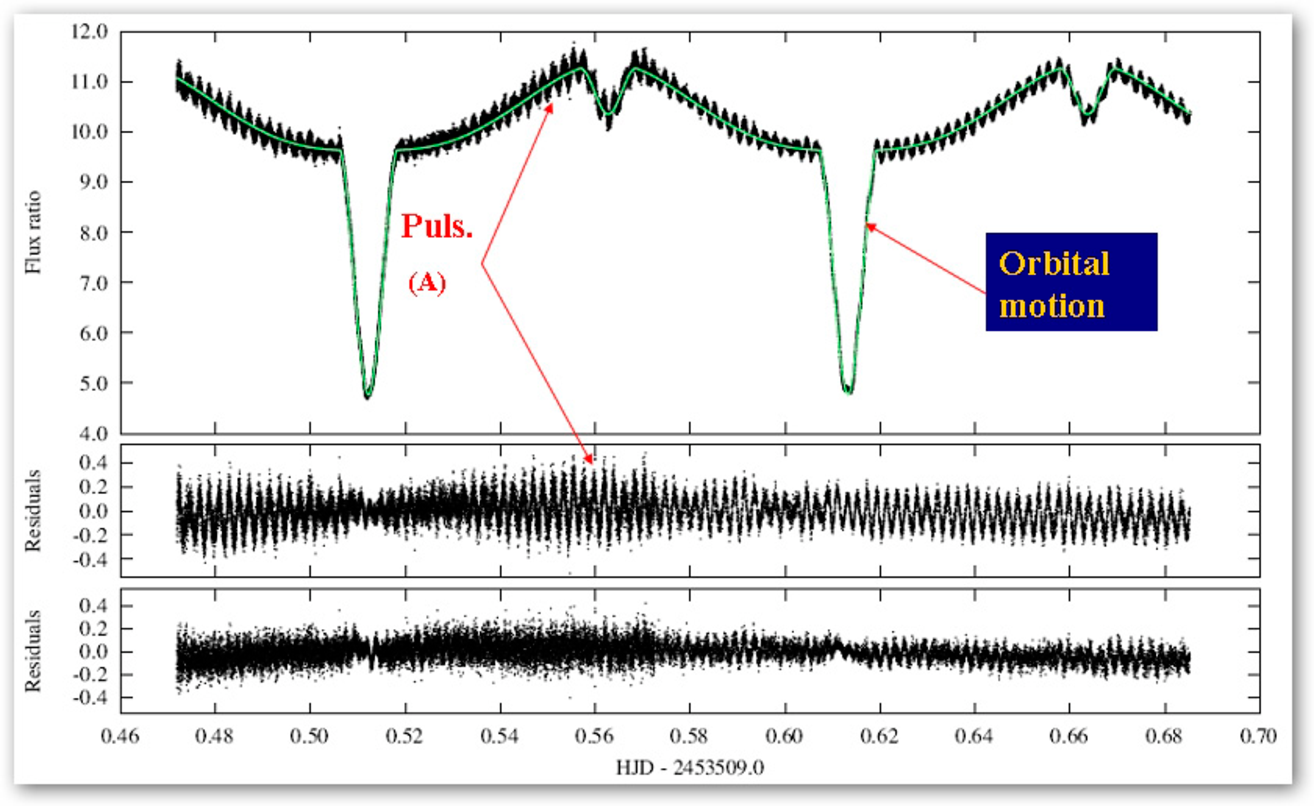}
\begin{minipage}[]{120mm}
   \caption[]{Sample light curves of PG 1336--018, an eclipsing binary with a pulsating subdwarf B component.
   Adopted from Vu$\check{c}$kovi\'{c} et al.(2007). }\end{minipage}
   \label{Fig:oEA-LC}
\end{figure}

What has caused the observational studies of eclipsing binaries (EB) with pulsating components
important?
Let us see the case of an oscillating Algol-type eclipsing binary system (designated
as oEA, following Kim et al.\,2003).
An oEA's light variations would contain that due to the reflect and proximity effects,
in addition to the eclipse,
while the remained periodic variations are intrinsic variability to
one of the components (usually the primary one) of the binary system.
Figure~\ref{Fig:oEA-LC} shows an example of such superimposed light variations.
The photometric analysis of such binaries is unavoidably affected by intrinsic variations
due to pulsation.
In some case, if the big-sized component is pulsating, then periodic intrinsic pulsation
can be seen during the whole orbital period, even in full eclipse.
Most oEA stars exhibit the pulsational properties of a typical $\delta$ Scuti star.
However, their evolutionary history is entirely different with respect to single pulsators.
This evokes not only one's observational interests
but it also brings about a challenge to both stellar evolutionary and pulsational
theories in characterizing the pulsators and binary systems.
Eclipsing binaries, as one of the fundamental astrophysical objects,
when coupled with asteroseismology, they provide two independent
methods to obtain masses and radii and excellent opportunities to develop
highly constrained stellar models.

Research suggests that many stars are part of either binary star systems
or star systems with more than two stars, called multiple stellar systems.
Some 67\% of the stars with spectral types ranged G to M in the vicinity of the Sun are
binaries (Mayor et al. 2001). Moreover, 75\% high-mass binaries are found in star clusters
and stellar associations (of O--B types, Mason et al.\,2001).

\begin{figure}[t!!!]
   \vspace{2mm}
   \centering
   \includegraphics[width=146mm, height=110mm,angle=0,scale=1.0]{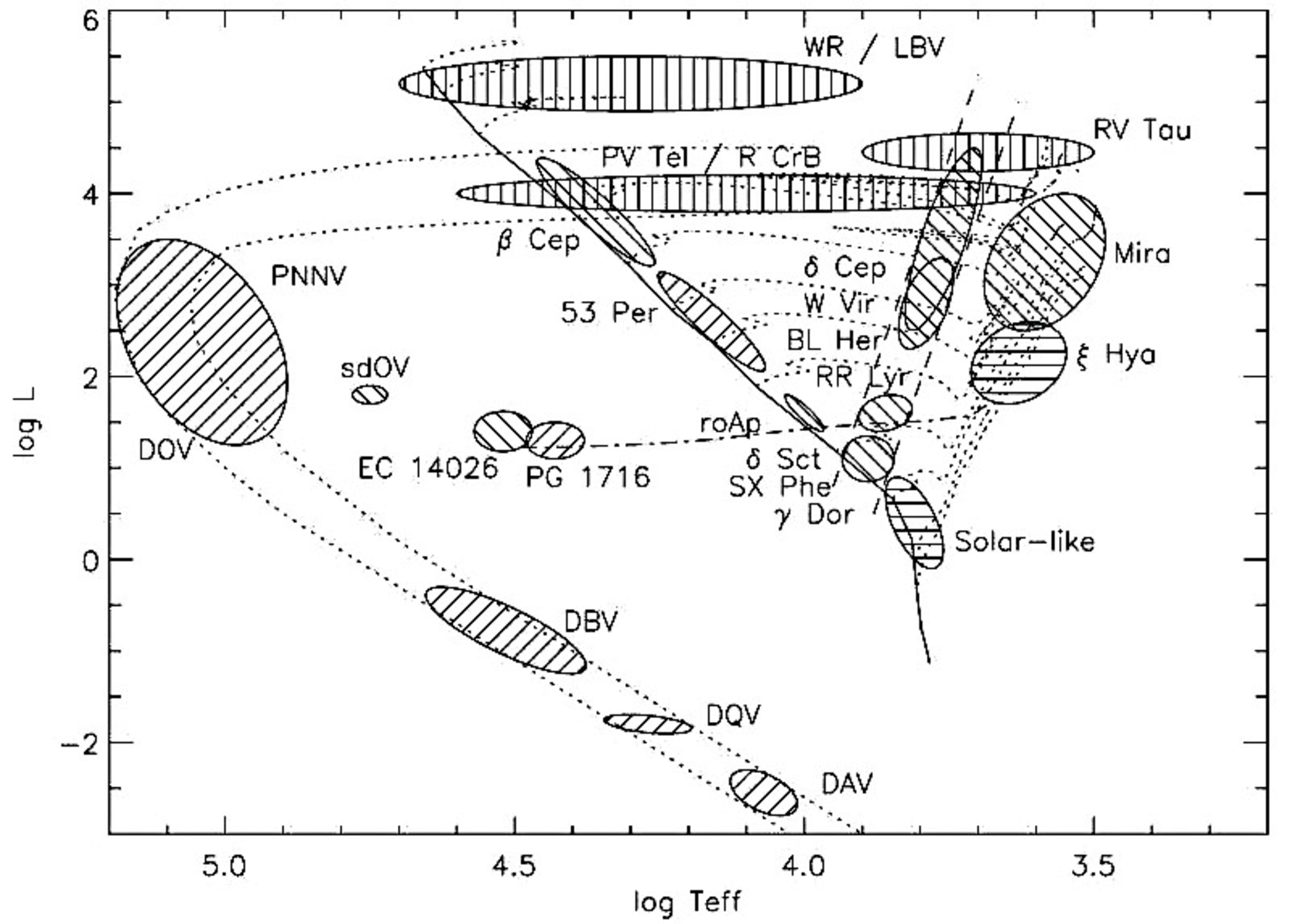}
\begin{minipage}[]{145mm}
   \caption{A version of the Hertzsprung-Russell diagram of pulsating stars.
   Adopted from Jeffery (2008).}\end{minipage}
   \label{Fig:HR-diagram}
\end{figure}

Pulsation has been detected in stars of almost every kind of spectral types.
Pulsating variables cover a wide range of masses and almost every stage in
stellar evolution.
Pulsating variables and binary systems are connected with each other over a long history.
Almost all types of pulsating stars (see Fig.\,\ref{Fig:HR-diagram} or
its original version, fig.1 of Jeffery 2008) are found as members of binary systems.
For example,
nearly a half of the 60 spectroscopically monitored $\gamma$ Doradus stars in
northern sky are binaries (Mathias et al.\,2004);
the majority of classical Cepheids have one or more companions (Szabados 2003b).

The period-luminosity relationship of Cepheids makes their study one of
the most effective ways to measure the distance to nearby galaxies and
thus to map out the scale of the whole universe.
This useful feature of Cepheids has earned them the nickname ``standard candles".
Unfortunately, despite their importance for the improvement in the cosmic distance scale,
Cepheids are not fully understood.
Predictions of their masses derived from the theory of pulsating stars are
20\% less than predictions from the theory of stellar evolution.
The find of eclipsing binaries containing a Cepheid will help in accurately measuring
the orbital motion, sizes and masses of the two stars.
It is the best approach to solving the above mass discrepancy.
Unfortunately neither Cepheids nor eclipsing binaries are common,
so the chance of finding such an unusual pair seemed incredibly rare.
None are known in the Milky Way by now.
MACHO 81.8997.87 was first identified as an eclipsing Cepheid (in first overtone mode)
in the Large Magellanic Cloud (LMC) by OGLE (Udalski et al. 1999),
and it is reconfirmed by MACHO (Lepischak et al. 2004).
Eclipsing binary systems with Cepheid components in the LMC is
a key to the extragalactic distance scale (Guinan et al. 2005).
The discovery of the eclipsing double star
OGLE-LMC-CEP 0227 in the LMC (Pietrzy\'{n}ski et al.\,2010),
where a 3\fd8-pulsating Cepheid variable orbiting another star in a period of 310 days.
The rare alignment of the orbits of the two stars in this eclipsing system
has allowed a measurement of the Cepheid mass with unprecedented accuracy.
The new result shows that the prediction from stellar pulsation theory is spot on,
while the prediction from stellar evolution theory is at odds with the new observations.

Similar to Cepheids,
because of their constant mean luminosities $\sim 45\,L_{\sun}$ ($\langle M_V \rangle \sim$ +0.5 mag)
and easily recognizable light curves,
RR Lyr variables have long served as the ``cornerstone" of
the Pop. II distance scale in our galaxy and for local group galaxies.
However, there are few fundamental data available for RR Lyr stars.
In fact, except the trigonometric parallax determination of RR Lyr itself from
HST measurements (Benedict et al.\,2002),
there are few direct measures of their most fundamental properties
--- such as mass, radius and luminosity.
The astrophysical and cosmological consequences of finding an RR Lyr star in
an EB are considerable, because the masses and absolute radii of the components of
eclipsing binaries can be determined to within a few percent from
analyses of their light and radial velocity curves.
RW Ari was early suspected to be in an EB system (Wi\'{s}niewski 1971)
but it was not supported by late observations (e.g. Dahm 1994).
The RR Lyr star TU UMa was highly suspected to be a member
of a binary system by Wade et al.(1999), but without follow-up confirmation.
OGLE data could have discovered three candidates in the LMC which
simultaneously reveal RR Lyr-type and eclipsing-type variability (Soszy\'{n}ski et al.\,2003).
One of the three candidates OGLE J052218.07-692827.4 apparently showed
detached eclipsing binary signature.
The RR Lyr primary component has a pulsation period of 0.564876 days.
It was suspected to be the only bona fide EB with
an RR Lyr component (Guinan et al.\,2007).
However, HST/WFPC2 observations of the star resolved 5 distinct sources
within a 1.3$^{\prime\prime}$ region --- the typical OGLE resolution,
proving that it is also an optical blend,
does not seem to correspond to a physically plausible system.
The source is likely another background RR Lyr star.
As of 2008, there is still no an RR Lyr star discovered in
an eclipsing binary system (Pr$\check{\rm s}$a et al.\,2008).
Most recently, using the OGLE-III database, Soszy\'{n}ski et al.(2011) reported
the breakthrough discovery of an RR Lyr star OGLE-BLG-RRLYR-02792 with additional
eclipsing variability with the orbital period of 15.2447 days.
These authors have identified further three RR Lyr stars being likely in EB systems.

Recent discoveries of various oscillating stars in eclipsing binaries have motivated the researchers,
who are specialized in the fields of binaries and pulsating variables,
to draw their observational attention onto oscillating binary systems.
For example,
the discovery of non-radial pulsations in the Herbig Ae type spectroscopic binary
RS Cha (B\"{o}hm et al.\,2009);
the discovery of slow X-ray pulsations in the high-mass X-ray binary 4U 2206+54
(Reig et al.\,2009);
the search for planets around pulsating subdwarf B stars (Schuh et al.\,2010);
the detection of a tertiary brown dwarf companion in the sdB-type eclipsing binary HS 0705+6700
(Qian et al.\,2009) and circumbinary planets orbiting the sdB binary NY Vir (Qian et al. 2012);
the detection of a giant extrasolar planet orbiting the eclipsing polar DP Leo
(Qian et al.\,2010), which is identified as a cataclysmic binary by Beuermann et al.(2011).

Most recently, a search for radio pulsations from neutron star companions of
four subdwarf B stars (Coenen et al.\,2011) leads to the results that they orbit
a companion in the neutron star mass range.
Such companions are thought to play an important role in the poorly understood
formation of subdwarf B stars.

\section{Rationale and Benefit of Studying Binary Systems Oscillating Components}
There are only two underlying mechanisms for driving stellar pulsations:
the self excitation in the layers (which operate as a heat engine)
and the stochastic oscillations by turbulent convection.
The former instability mechanism excites pulsations in most stars, beginning from
classical instability strip stars, through B type main sequence stars,
hot subdwarfs to white dwarfs.
The second stochastic excitation drives solar-like oscillations,
including those observed in the Sun, and is expected in all stars with
extended convective outer layers (Daszy\'{n}ska-Daszkiewicz\,2009).
That is not the whole story of stellar pulsation.
In pulsation modelling, pulsation modes are needed to be observationally identified first.
More importantly, the stellar mass must be well determined as a key input parameter
before applying asteroseismic techniques.

How can we measure the mass of a distant star?
As known, stellar mass (together with radius) can be independently and uniquely determined
only if the star is a component of a double-lined spectroscopic eclipsing binary
by the Kepler's third law:
\begin{equation}
\frac{G(M_1+M_2)}{a^3} = \frac{4\pi^2}{P^2}  {~~~~~~\rm or~~~~~~}  M_1+M_2 = \frac{a^3}{P^2} ~,
\end{equation}
if one measures orbital period $P$ in years,
semimajor axis of the orbit $a$ (or separation of the two components)
in astronomical units (AU) and each component's mass ($M_1, M_2$) in solar masses.
The orbital velocity amplitudes ($K_1, K_2$) are used to determine $a$ directly
by the relation
\begin{equation}
|K_1|+|K_2|=\frac{2\pi}{P} a\sin i~.
\end{equation}
A general mass function for the secondary companion can be written as
\begin{equation}
M_2\sin i = (\frac{P}{2\pi G})^{1/3} \frac{K_1 \sin i}{(M_1+M_2)^{2/3}} ~,
\end{equation}
where $K_1 \sin i$ is the line-of-sight velocity component of the orbital motion of
the primary companion about the center of mass, $G$ is the universal gravitational constant.
If the orbital plane of a binary system is perpendicular to the plane of the sky,
that is, we are observing in the orbital plane (edge-on, orbital inclination
$i=90\dg$, which is true if the system is eclipsing),
then the solution to the above equation is straightforward.
By using eq.(3), orbital eccentricity is neglected and circular orbit has been assumed.
Thus if an eclipsing binary is a double-lined spectroscopic system,
the fundamental physical properties (mass, radius, temperature and luminosity) can
be directly determined from the analysis of
the combined radial velocity and photometric observations.
The parameter relations given in eqs.(1)--(3)
is one of the main concerns on the study of eclipsing binary systems with pulsating components.

To summarize, the common interests in studying the pulsators in binary and multiple stellar systems
relies at least on following advantages:\\
\indent(1) If no mechanism damping pulsations in close binaries,
percentage of pulsating stars expected among A--F components of
detached and semi-detached eclipsing binaries should be at least
the same as for single A--F type stars.\\
\indent(2) Pulsation characteristics of oscillating binaries are similar to those of
single pulsators, but their evolutions are quite different due to mass accretion.\\
\indent(3) Precise estimation of the accretion rate using the pulsation period changes
of the gainer(i.e. accreting star) caused by accretion.\\
\indent(4) Possibility of precise dynamical mass and radius determinations --- the masses and radii for
each component in eclipsing (double-lined) spectroscopic binaries
could be accurately determined. \\
\indent(5) With a certain mass, it should help to model the pulsating spectra.\\
\indent(6) The possibility of (non-radial) mode identification during the eclipse orbital phases
(i.e. the primary minima) using the observed pulsational amplitude and
phase changes during the eclipse (Nather \& Robinson\,1974).
This has been explored by, for instance, Reed et al.\,(2005) and B\'{\i}r\'{o} \& Nuspl\,(2011).\\
\indent(7) Higher probability of detection of the sectorial modes due to equator-on visibility
of components in close eclipsing binaries.\\
\indent(8) When applying asteroseismic diagnostic tools to studying the dynamics of mass transfer
between components in semi-detached eclipsing binaries, the possibility of precise estimation of
the accretion rate would become higher if using the pulsation period changes of
the gainer star caused by accretion.\\
\indent(9) The study of pulsational properties of the pulsating components in eclipsing binaries
is in its blossom state, whereas asteroseismology of these stars is very attractive
in comparison to single stars --- pulsational properties can be constrained using
spectroscopic eclipsing binary systems (say Creevey et al. 2011).\\
\indent(10) The connection between asteroseismology and exoplanet research,
i.e. the study of pulsating stars harboring planets. Two early discussions can be found in
Moya (2013) and Vauclair (2008, 2012). A recent technique progress using photometric data to
derive radial velocity is presented by Shibahashi \& Kurtz (2012) and Murphy et al.(2014).
This technique has opened a new era in the EB study.\\
\indent(11) Eclipsing binaries containing pulsating stars provide a unique opportunity to improve
calibration of the cosmic distance scale and to better calibrate stellar evolutionary models.

By now, more than ten types of pulsating stars are found as members of various binary systems.
Those planet-hosting oscillators can also be regarded as a special case of binary systems.
At the time of observational efforts rolling into this field,
the binary, triple or multiple stellar systems with pulsating components are needed
to be collected in a catalog.
We have attempted to catalog both the Galactic and extragalactic stellar systems
with pulsating components.
However, the number of pulsating binaries is increasing.
As of this writing, some oscillating multiple stellar systems
might have not been collected.
The readers should be aware that other kinds of
pulsating binary systems not mentioned in current version are possible.
Regarding this, the missing materials will be supplied in future updates.

\section{Notes to the Catalog}
\label{sect:notes}

\setcounter{table}{0}
\begin{table*}[t!!!]
\begin{minipage}[]{70cm}
\caption[c]{~~A statistics on pulsating binaries based on the presented catalog.
Candidates follow the plus '+' sign.}
\end{minipage}
   \label{Tab:stat}
   \vspace{-3mm}
   \begin{center}
   \begin{tabular}{lccccc}
\hline\noalign{\smallskip}
Type of Pulsators  & & \multicolumn{4}{c}{Classes of stellar systems} \\
                    \cline{3-6}
                                     &Group     & Eclipsing & Spectroscopic & Visual  & Others$^{\dag}$\\
                                     &Sum       & Binaries  & Binaries      & Binaries& \\
   \hline\noalign{\smallskip}
~~(1)~ DCEP: Galactic$^{\dag\dag}$   & 154 + 34 &    3      & 123           & 20      & 20 \\ 
~~~~~~~ DCEP: Extragalactic          & ~~4 + ~0 &    4      &   0           &  0      & -- \\
~~~~~~~ Type II Cepheids(CWA,CWB,RV):& ~14 + 18 &    7      &   7           &  0      & -- \\
~~(2)~ DSCT: $\delta$ Scuti-type     & 112 + 53 &   96      &  13           &  1      &  7 \\
~~(3)~ solar-like oscillators + RGB  & ~71 + ~0 &   65      &   4           &  2      &  1 \\
~~(4)~ sdBV: pulsating subdwarf B/O  & ~41 + 20 &   15      &  24           & --      &  1 \\
~~(5)~ CV: cataclysmic variable      & ~32 + 28 &   29      &   3           &  0      &  0 \\
~~(6)~ BCEP: $\beta$ Cep-type        & ~26 + ~9 &   10      &  11           &  --     &  5 \\
~~(7)~ SPB: slowly pulsating B stars & ~15 + ~0 &    3      &  10           &  2      &  2 \\
~~(8)~ GDOR: $\gamma$ Dor-type       & ~12 + 30 &   10      &   2           &  0      &  -- \\
~~(9)~ Be/X-ray pulsators            & ~10 + ~0 &    9      &   0           &  0      &  -- \\
(10)~ WD: pulsating white dwarf      & ~~6 + 57 &    3      &   3           &  0      &  -- \\
(11)~ WR: Wolf-Rayet stars           & ~~3 + ~0 &    3      &   0           &  0      &  -- \\
(12)~ SX Phe-type                    & ~~6 + ~3 &    2      &   0           &  0      &  -- \\
(13)~ BY Dra-type                    & ~~1 + ~0 &    1      &   0           &  0      &  -- \\
(14)~ HADS: high-amplitude DSCT      & ~~2 + ~0 &    1      &   0           &  0      &  -- \\
(15)~ RR: RR Lyr-type                & ~~4 + 59 &    1      &   0           &  0      &  -- \\
(16)~ non-classified                 & ~~2 + 11 &    1      &   1           &  0      &  -- \\
\hline
Total                                & 515~+~322$^*$&  262      & 201           & 23      & 36  \\
   \hline\noalign{\smallskip}
   \end{tabular}
   \end{center}
\vspace{-3mm}\hspace{16mm} \dag: Column `Others'  for triple/multiple systems and unidentified;\\
\vspace{-0.0mm}\hspace{16mm}\dag\dag: Galactic classical Cepheids in binaries are adopted from Szabados (2003a);\\
\vspace{-0.0mm}\hspace{16mm} $^*$: These candidates were extracted from Simbad database without literature check for their identities.
\end{table*}

\subsection{General notes}

Labels used in the catalog:\\
\indent (1)~ Sp.(A+B) --- spectral types of the primary (A) and secondary component (B)\\
\indent (2)~ $\langle V \rangle$      --- mean magnitude in $V$ band\\
\indent (3)~ $\langle B \rangle$      --- mean magnitude in $B$ band\\
\indent (4)~ $P_{\rm orb}$ --- orbital period in days\\
\indent (5)~ $P_{\rm pul}$ --- main pulsation period in days (except those indicated units)\\
\indent (6)~ Comments --- key characteristics of the pulsating multiple stellar systems,
membership of a cluster, other identifications, etc.
Full identification in the survey project is provided, some object names used
short nomenclature following the original references.\\
\indent (7)~ EB; EA --- Eclipsing binary (system); Eclipsing binary of Algol-type\\
\indent (8)~ oEA --- oscillating EA (eclipsing binaries of Algol-type)\\
\indent (9)~ SB --- refers to spectroscopic binary: SB1 (single-lined), SB2 (double-lined).\\
\indent (10) SB+orbit --- stands for spectroscopic binary with known orbital elements available in literature.\\
\indent (11) comp.? --- photometric companion, physical relation should be investigated.\\
\indent (12) References --- key references related to the pulsational properties and binarity.
Some data were adopted from Soydugan et al.\,(2006a), which is not always listed
when original papers or latest results are cited.
Columns missing data will be populated in the future updates.\\

\subsection{Comments on each subgroup of the pulsating multiple stellar systems}

\begin{figure}[t!!!]
   \vspace{-12mm}\centering
   \includegraphics[width=146mm, height=110mm,angle=0,scale=1.0]{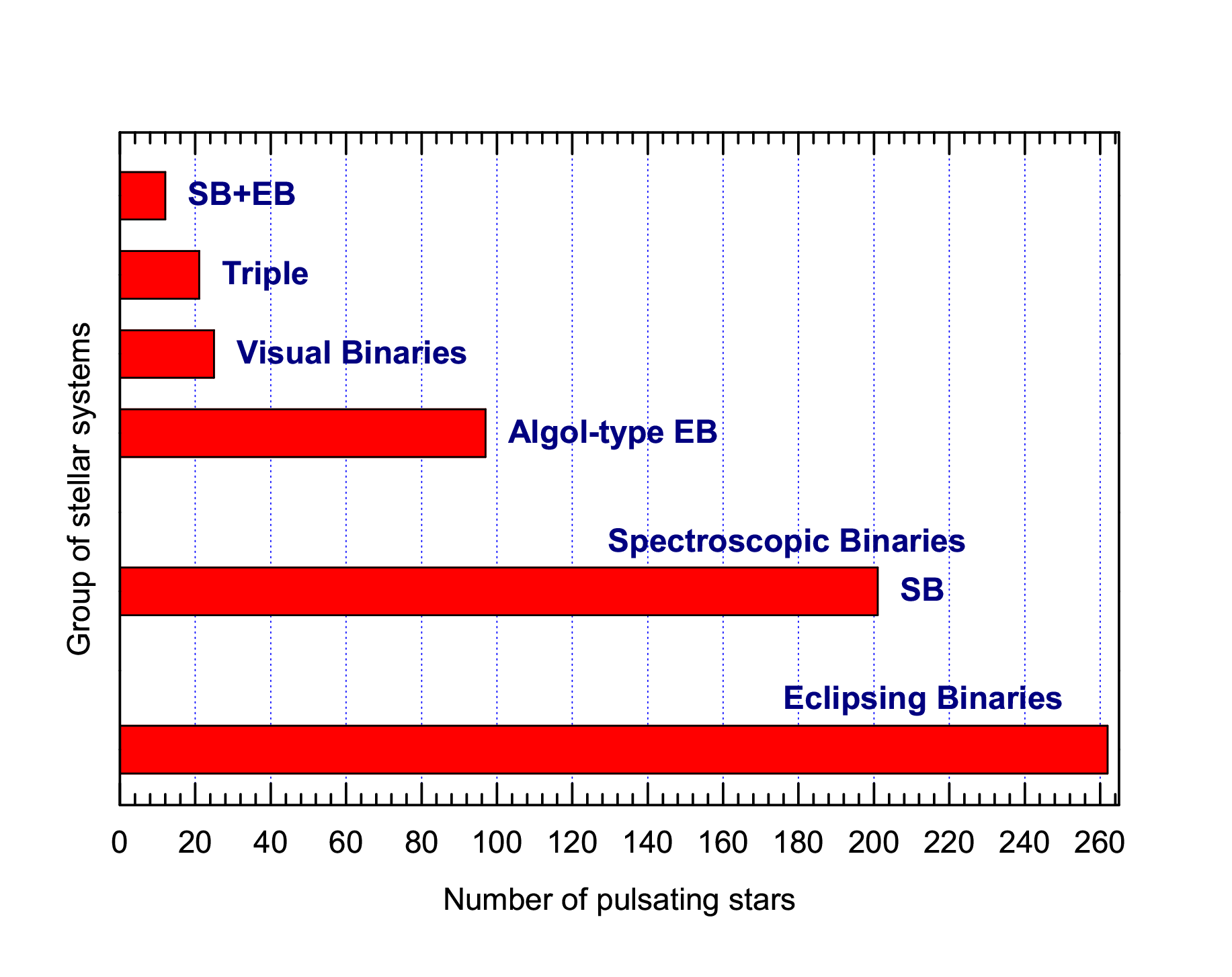}

\begin{minipage}[]{120mm}
   \vspace{-12mm}
   \caption{Number of pulsators in different groups of multiple stellar systems.}
\end{minipage}
   \label{Fig:number}
\end{figure}

\begin{figure}[th]
   \vspace{-12mm}\centering
   \includegraphics[width=150mm, height=110mm,angle=0,scale=1.0]{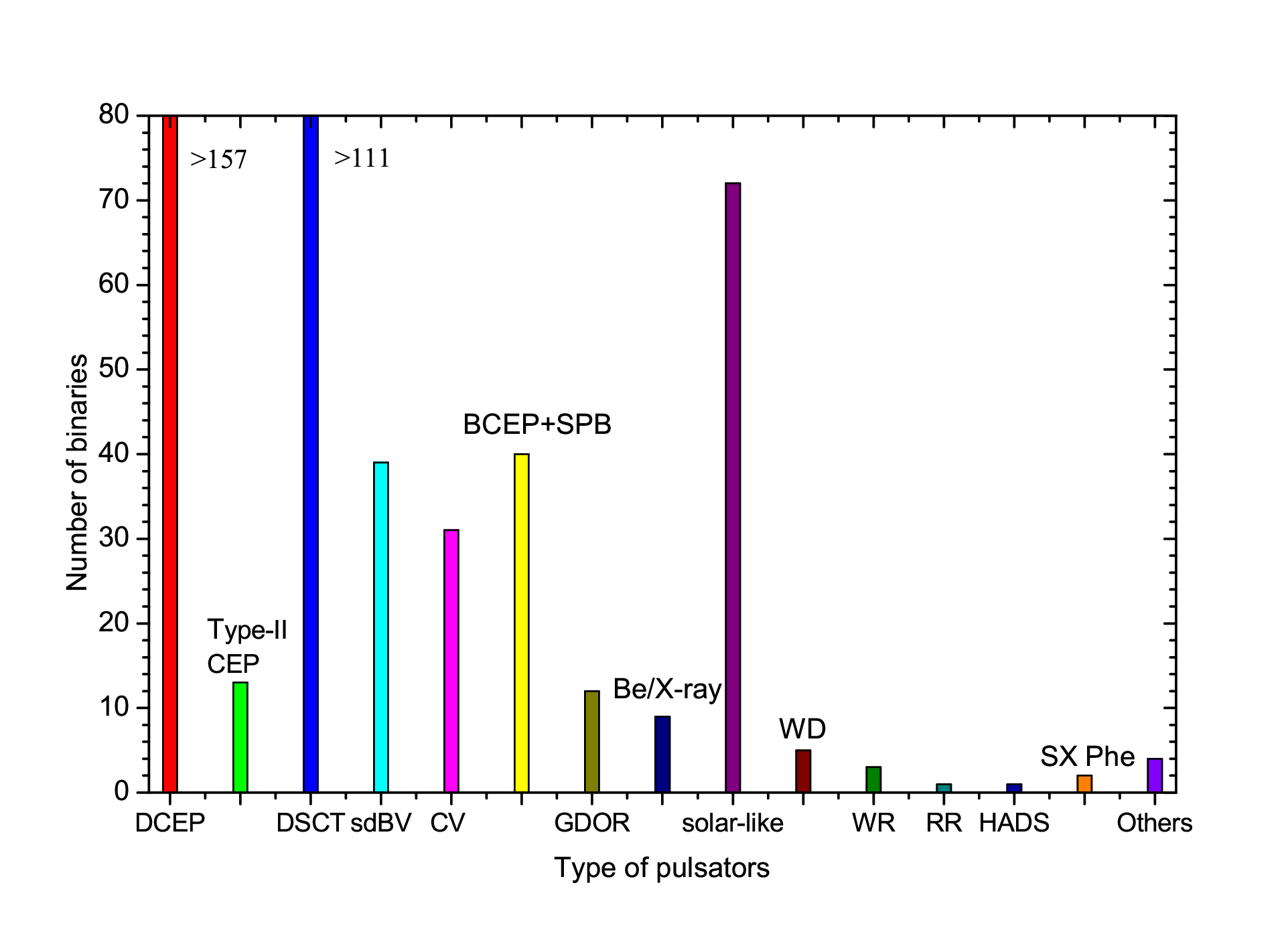}
   \vspace{-10mm}
   \caption{Distribution of binaries (and multiples) containing different type pulsators.
   The columns for Galactic DCEP and $\delta$ Sct are cutoff in order to display others better.    }
   \label{Fig:dist}
\end{figure}

Under the scope of pulsating components in binaries,
we have summarized, as a catalog in Table\,\ref{Tab:stat}, all the types of
pulsating stars currently discovered in various groups of binary systems.
We describe each type of pulsating stars and binary systems briefly
according to the GCVS as a conceptual background knowledge.

\subsubsection{Types of Binary Systems Involved}

\begin{enumerate}
  \item EB:~ Eclipsing binary (systems). These are binary systems with orbital planes
       so close to the observer's line of sight (the inclination of the
       orbital plane to the plane orthogonal to the line of sight is
       close to $90\dg$) that the components periodically eclipse each other.
       Consequently, the observer finds changes of the apparent combined
       brightness of the system with the period coincident with that of the
       components' orbital motion.

  \item EA:~ eclipsing binaries of Algol ($\beta$ Persei)-type,
       are binaries with spherical or slightly ellipsoidal (usually semi-detached or detached) components.
       It is possible to specify, for their light curves, the moments of the beginning and
       end of the eclipses.
       Between eclipses the light remains almost constant or varies insignificantly
       because of reflection effects, slight ellipsoidality of components,
       or physical variations. Secondary minima may be absent.
       An extremely wide range of periods is observed,
       from 0.1167 (HW Vir) to $\geq$ 10\,000 days ($\epsilon$ Aur).
       Light amplitudes are also quite different and may reach several magnitudes.
       For the oscillating Algol-type eclipsing binaries (widely recognized as oEA systems),
       usually the primary components are intrinsic pulsating variable
       stars (e.g. $\delta$ Scuti-type pulsators),
       while the late-type secondary fills its Roche lobe.
       Distribution of a sample of 434 confirmed EA with respect to spectral types is given in
       Fig.~\ref{Fig:EA-dist}. These binaries are provided in appendix.

  \item EB($\beta$):~ $\beta$ Lyrae-type eclipsing systems. These are eclipsing systems having
       ellipsoidal components and light curves for which it is impossible
       to specify the exact times of onset and end of eclipses because of
       a continuous change of a system's apparent combined brightness
       between eclipses; secondary minimum is observed in all cases, its
       depth usually being considerably smaller than that of the primary
       minimum; periods are mainly longer than 1 day. The components
       generally belong to early spectral types (B--A). Light amplitudes
       are usually $<2$ mag in $V$.

  \item EW:~ W Ursae Majoris-type eclipsing variables (W UMa-type). These are eclipsing binaries with
       periods shorter than 1 day, consisting of ellipsoidal components
       almost in contact (some even overcontact) and having light curves for which it is
       impossible to specify the exact times of onset and end of
       eclipses. The depths of the primary and secondary minima are
       almost equal or differ insignificantly. Light amplitudes are
       usually $<$0.8 mag in $V$. The components generally belong to
       spectral types F--G and later.
       Only eight pulsating components are identified in EW by now.
However, several additional candidates were suggested (e.g. Michalska \& Pigulski 2008).

  \item SB:~ single- or double-lined spectroscopic binaries (SB1, SB2).

  \item X:~  Close binary systems that are sources of strong, variable X-ray
       emission and which do not belong to or are not yet attributed to any
       of the above types of variable stars. One of the components of
       the system is a hot compact object (white dwarf, neutron star, or
       possibly a black hole). X-ray emission originates from the infall
       of matter onto the compact object or onto an accretion disk
       surrounding the compact object. In turn, the X-ray emission is
       incident upon the atmosphere of the cooler companion of the
       compact object and is re-radiated in the form of optical
       high-temperature radiation (reflection effect), thus making that
       area of the cooler companion's surface an earlier spectral type.
       These effects lead to quite a peculiar complex character of
       optical variability in such systems.
       Be/X-ray pulsating binary systems is a type of this class.
\end{enumerate}

\subsubsection{$\delta$ Sct Pulsators in Binaries}
   Variables of the $\delta$ Scuti type (DSCT) are pulsating variables of
       spectral types A0-F5 with luminosity classes V to III
       displaying light amplitudes from 0.003
       to 0.9 mag in $V$ band (usually several hundredths of a magnitude) and
       periods from 0.01 to 0.2 days. The shapes of the light curves,
       periods, and amplitudes usually vary greatly. Radial as well as
       nonradial pulsations are observed. The variability of some
       members of this type appears sporadically and sometimes completely
       ceases, this being a consequence of strong amplitude modulation
       with the lower value of the amplitude not exceeding 0.001 mag
       in some cases. The maximum of the surface layer expansion does not
       lag behind the maximum light for more than 0.1 periods. DSCT stars are
       representatives of the galactic disk and are
       phenomenologically close to the SX Phe variables.
       They pulsate in radial and nonradial $p$ (pressure, and possibly also $g$ -- gravity) modes.
       After white dwarfs, they are the second most abundant pulsating variables in our Galaxy.

$\delta$ Sct type pulsators are driven by
the so-called $\kappa$ mechanism.  These stars pulsate mostly in low-radial-order pressure-mode
with pulsation constants usually less than 0.03\,d.
Seeds (1972) argued that about one third of 155 $\delta$ Sct stars are binary,
but as of 1974, only two (AB Cas and Y Cam) were known in eclipsing binaries.
Fitch (1976) suggested that high-amplitude DSCT are single,
while the low-amplitude ones probably have companions.
However, probably due to difficulty in observing small-amplitude oscillations in comparison with
large light variation caused by the eclipsing phenomenon,
only nine $\delta$ Sct-type pulsating components in EB systems were reported
as of 2001 (Rodr\'{\i}guez \& Breger 2001). Nevertheless, with high-precision CCD photometry and
various surveys including the space missions {\it CoRoT} and {\it Kepler},
the number has been inspiringly increased largely to more than 90 in recent years.

\begin{figure}[ht!!!]
   \vspace{-10mm}
   \centering
   \includegraphics[width=156mm, height=95mm,angle=0,scale=1.0]{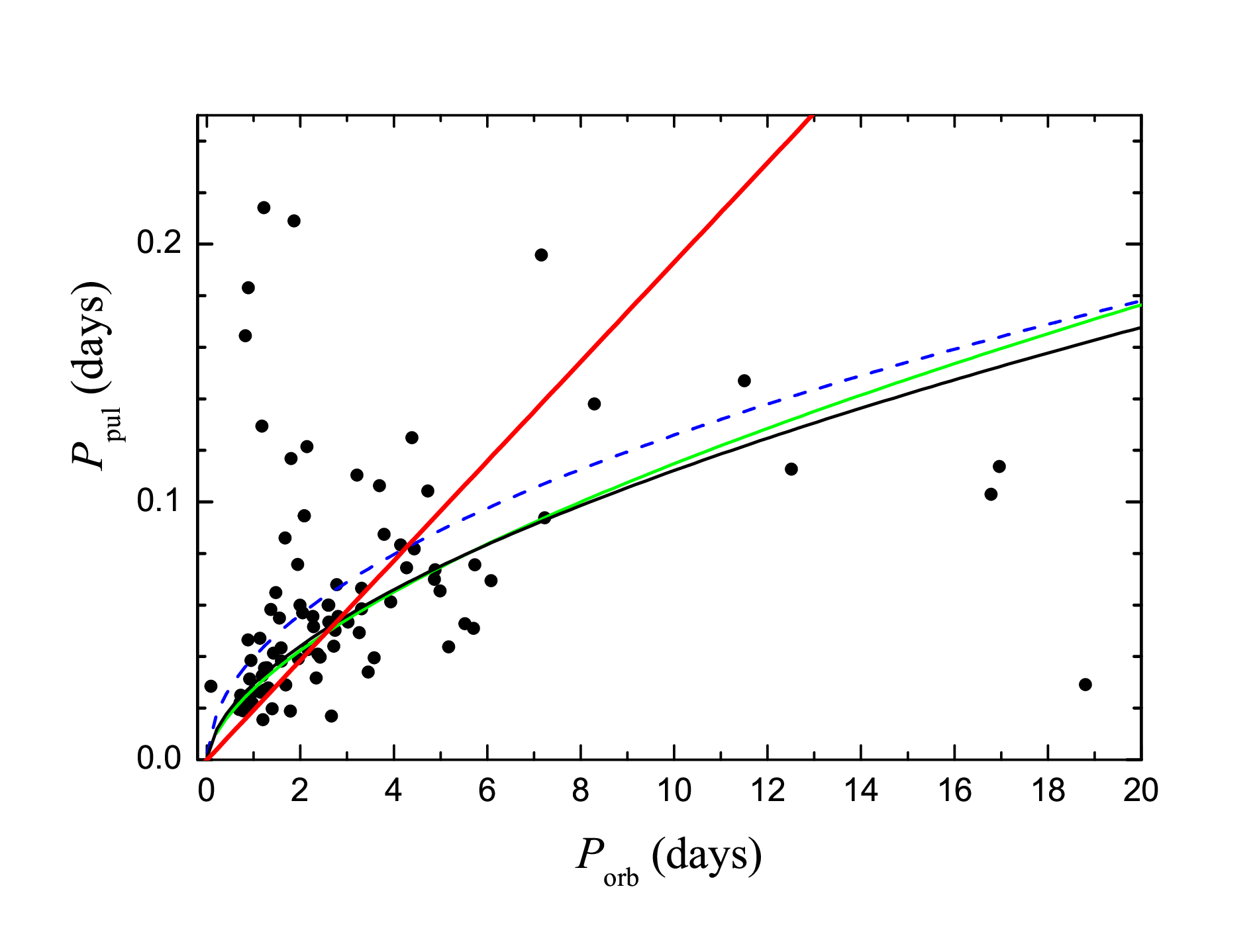}
   \includegraphics[width=156mm, height=90mm,angle=0,scale=1.0]{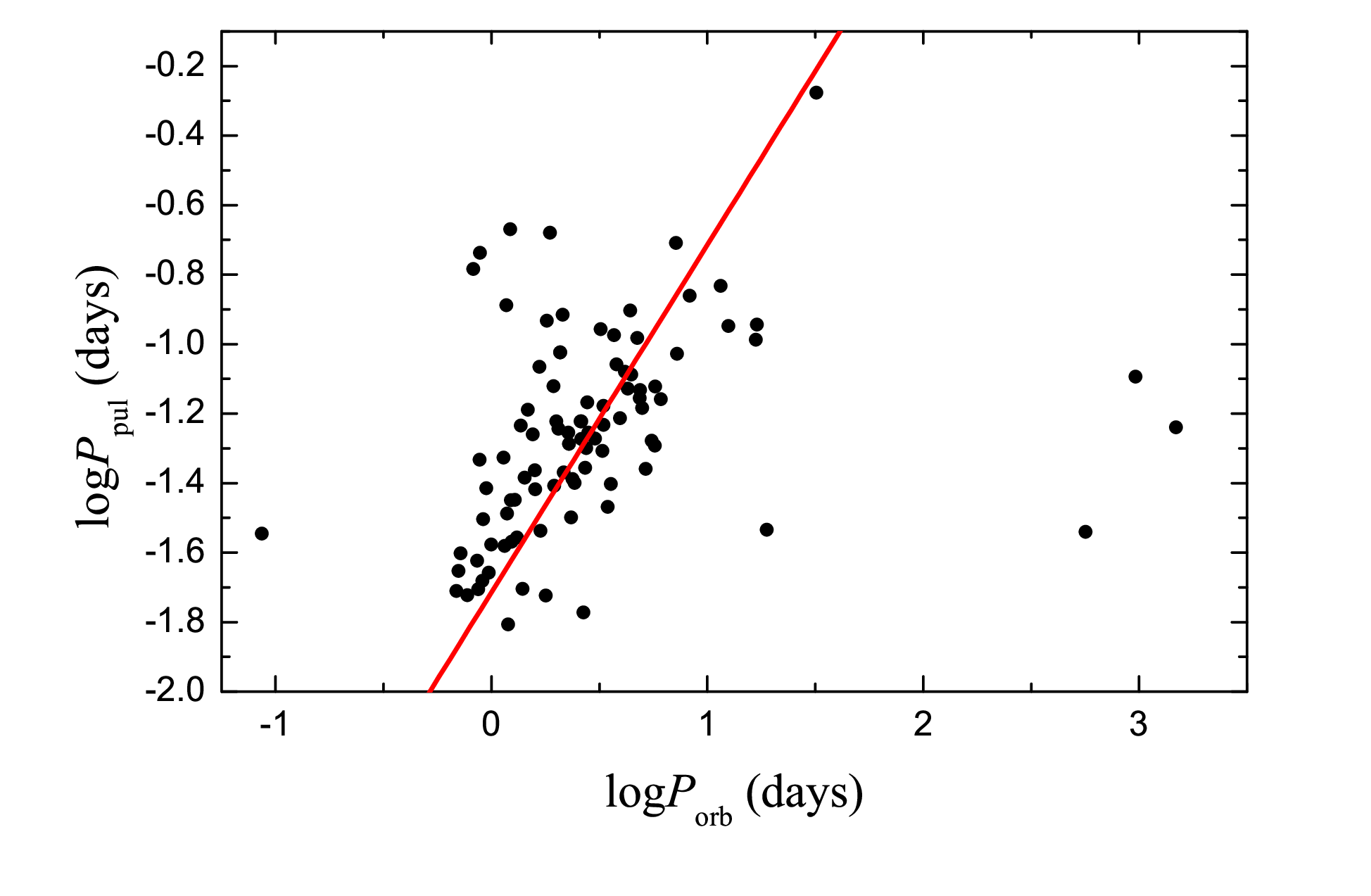}
   \vspace{-10mm}
   \caption{Correlation between the orbital and pulsational periods of
   the known 93 oEA with $\delta$ Sct components (DSCT-oEA).
   A few with longer orbital periods are cutoff.
   Red lines in both panels based upon a sample of 73 DSCT-oEA:
   $\log P_{\rm pul} = -1.7 + \log P_{\rm orb}$ (Zhang et al. 2013);
   Top:
   black solid line for a collection of 74 DSCT-oEA by Liakos et al.(2012): $\log P_{\rm pul} = -1.53 + 0.58\log P_{\rm orb}$ [eq.(3)];
   while the blue dash for semi-detached DSCT-oEA: $\log P_{\rm pul} = -1.4 + 0.52\log P_{\rm orb}$ [eq.(2)],
   green line for detached DSCT-oEA: $\log P_{\rm pul} = -1.56 + 0.62\log P_{\rm orb}$ [eq.(1)].
   }
   \label{Fig:oEA-orb-pul}
\end{figure}

\subsubsection{$\beta$ Cep and SPB Pulsators in Binaries}
There are two classes of B-type main sequence pulsators:
$\beta$ Cephei pulsators (BCEP) and slowly pulsating B stars (SPB).
Variables of the $\beta$ Cephei type (prototype $\beta$ Cep, $\beta$ CMa),
       are pulsating O8-B6 I-V stars with periods of light and
       radial-velocity variations in the range of 0.1--0.6 days and light
       amplitudes from 0.01 to 0.3 mag in $V$ band. The light curves are similar
       in shape to average radial-velocity curves but lag in phase by a
       quarter of the period, so that maximum brightness corresponds to
       maximum contraction, i.e., to minimum stellar radius. The
       majority of these stars probably show radial pulsations, but some
       (say V649 Per) display nonradial pulsations. Multiperiodicity is
       characteristic of many of these stars.
BCEP stars with masses larger than 8M$_{\sun}$ and spectral types B0--B2.5,
in which mainly pressure ($p$) modes are excited.

SPB stars are pulsating in high-order,
  low degree gravity ($g$) modes with typical periods of the order of a few days.
  These modes are excited by the opacity mechanism acting on the metal-bump.
  They are trapped deep in the interior of these hot stars, making them
  very interesting from an asteroseismic point of view. The theoretical
  pulsation frequency spectra of SPB stars are very dense, the observed
  amplitudes are low, and most of the currently known SPBs are multi-periodic,
  giving rise to beat periods of the order of months or even years.
  Rotation is a serious complication for mode identification in these stars because
  rotational splitting is large enough to cause multiplets of adjacent radial orders to overlap.
  Together with longer pulsation periods and rich eigen-spectra,
  great promise and obstacles coexist. These stars present serious observational
  and theoretical challenges.
  Currently, at least 51 confirmed and 65 candidate galactic SPB stars are
  known (Aerts et al.\,2006 and references therein),
  of which 15 are in open clusters.

  SPBs with spectral types B3--B9 and masses smaller than 8M$_{\sun}$.
  SPBs are somewhat similar to BCEP stars.
  Several BCEP/SPB hybrids are currently known,
  for example, $\gamma$ Peg (Handler 2009a,b),
  $\nu$ Eri and 12 Lac (Dziembowski \& Pamyatnykh 2008) have been confirmed to present
  both BCEP and SPB types of variations, i.e. pressure and gravity modes pulsation.
  The existence of excited $p$ and $g$ modes should allow
the simultaneous study of both the external and the internal zones of the stars.
It may also help to refine the limits of the SPBs instability zone.

\subsubsection{Cepheids in Binaries}
Cepheids (CEP) are radially pulsating, high luminosity (classes Ib-II)
  massive variables with periods in the range of 1--135 days and amplitudes from
       several hundredths to 2 mag in $V$ (in the $B$ band, the amplitudes
       are greater). Light curves show a rapid rise in brightness followed
       by a more gradual decline, shaped like a shark fin.
       Spectral type at maximum light is F; at minimum,
       the types are G--K. The longer the period of light variation,
       the later is the spectral type. The maximum of the surface-layer
       expansion velocity almost coinciding with maximum light.
       (1) Classical Cepheids (Pop.\,I, prototype $\delta$ Cephei)
       or called $\delta$ Cephei-type (DCEP) variables, are
       comparatively young objects that have left the main sequence and
       evolved into the instability strip of the Hertzsprung-Russell
       (H-R) diagram, they obey the well-known Cepheid period-luminosity
       relation and belong to the young disk population. DCEP stars are
       present in open clusters. They display a certain relation between
       the shapes of their light curves and their periods.
       (2) Type II Cepheids: metal-poor, low mass, population II.
       They are divided into three subclasses:
       BL Herculis-type (P=1--5 days), W Virginis-type (P=$\sim$10--20 days), and RV Tauri-type (P$>$20 days).
       W Vir Cepheids (CW) are pulsating variables of
       the galactic spherical component (old disk) population with
       periods of approximately 0.8 to 35 days and amplitudes from 0.3 to
       1.2 mag in $V$. They obey a period-luminosity relation different
       from that for DCEP variables.
       For an equal period value, the CW stars are fainter than the DCEP stars by
       0.7--2 mag. The light curves of CW variables for some period
       intervals differ from those of DCEP variables for
       corresponding periods either by amplitudes or by the presence of
       humps on their descending branches, sometimes turning into broad
       flat maxima. CW variables are present in globular clusters and
       at high galactic latitudes. DCEP and CW are distinct groups of entirely different
       objects in different evolutionary stages.
       A few RV Tauri-type pulsating variables were found in
       post-Asymptotic Giant Branch (post-AGB) binaries.

       During the last few years, observations have revealed that
       (1) nonradial modes in classical Cepheids (Moskalik \& Kolaczkowski 2008);
       (2) eclipsing binaries containing Cepheids (Soszy\'{n}ski et al. 2008b; Pietrzy\'{n}ski et al. 2010);
       (3) triple-mode Cepheids (Soszy\'{n}ski et al. 2008a), etc.
       Most stellar systems containing Cepheids are spectroscopic binaries,
       a few are eclipsing binaries. Moreover, the orbital period of
       these binaries usually are quite long (up to tens of years),
       see details in the present catalog.

The astrophysical and cosmological importance of finding a Cepheid as a member of
an eclipsing binary is considerable. If an eclipsing binary is a double-lined system,
the mass, radius, and luminosity can be directly determined from the analysis of
the light and radial velocity curves.
Moreover, the study of Cepheids in eclipsing binaries offers an important opportunity to
investigate the structure and evolution of Cepheids as well as tests of pulsational theories.
These systems provide opportunity to minimize the dependence of
the Cosmic Distance Scale and Hubble's constant on uncertainties
in assumed ``zero-points".
Unfortunately, there are no Cepheids in eclipsing binary systems known so far in the
Milky Way.

\begin{figure}[th]
   \vspace{2mm}
   \centering
\vspace{-6mm}
   \includegraphics[width=130mm, height=85mm,angle=0,scale=1.0]  {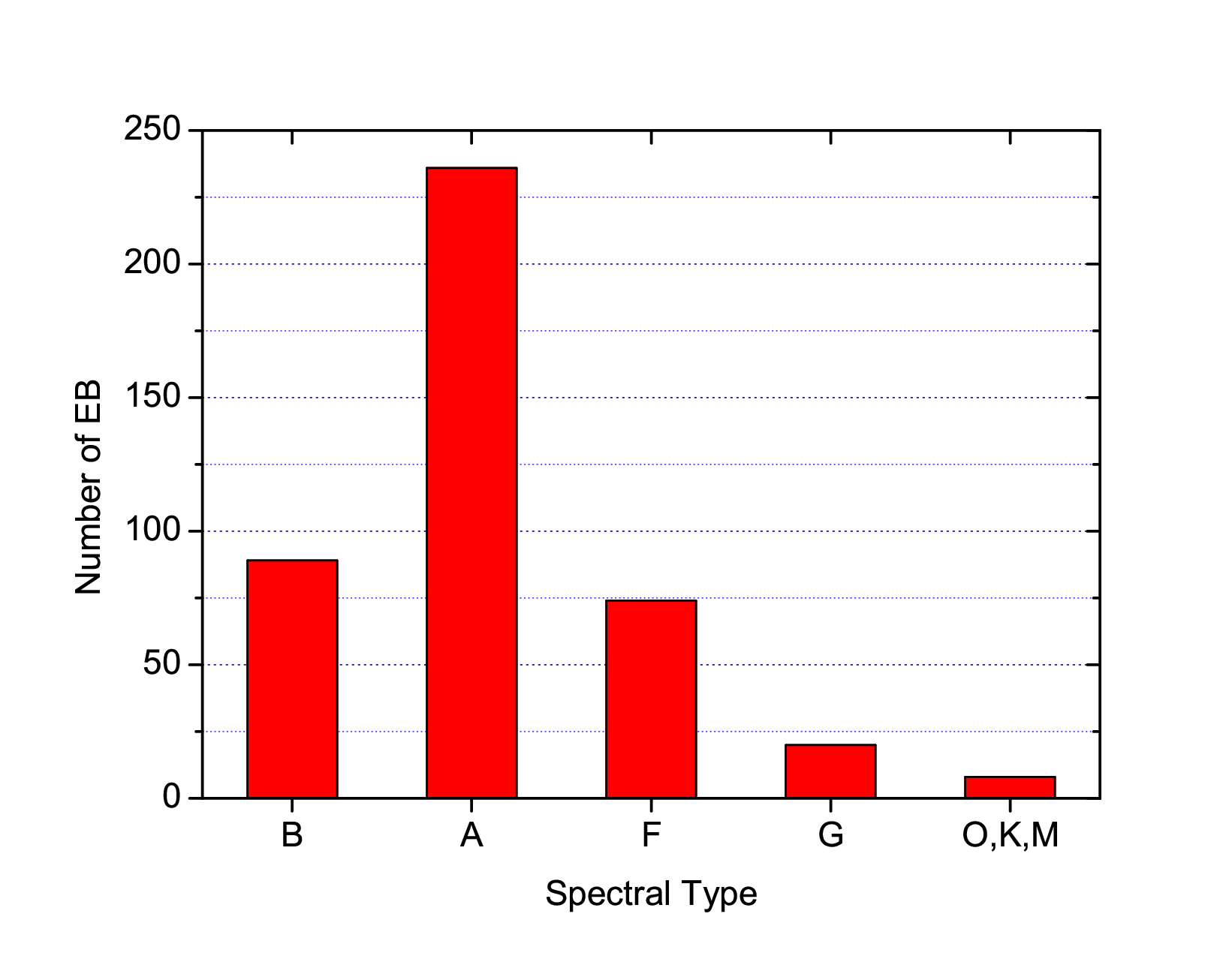}
   \includegraphics[width=80mm, height=145mm,angle=-90,scale=1.0]{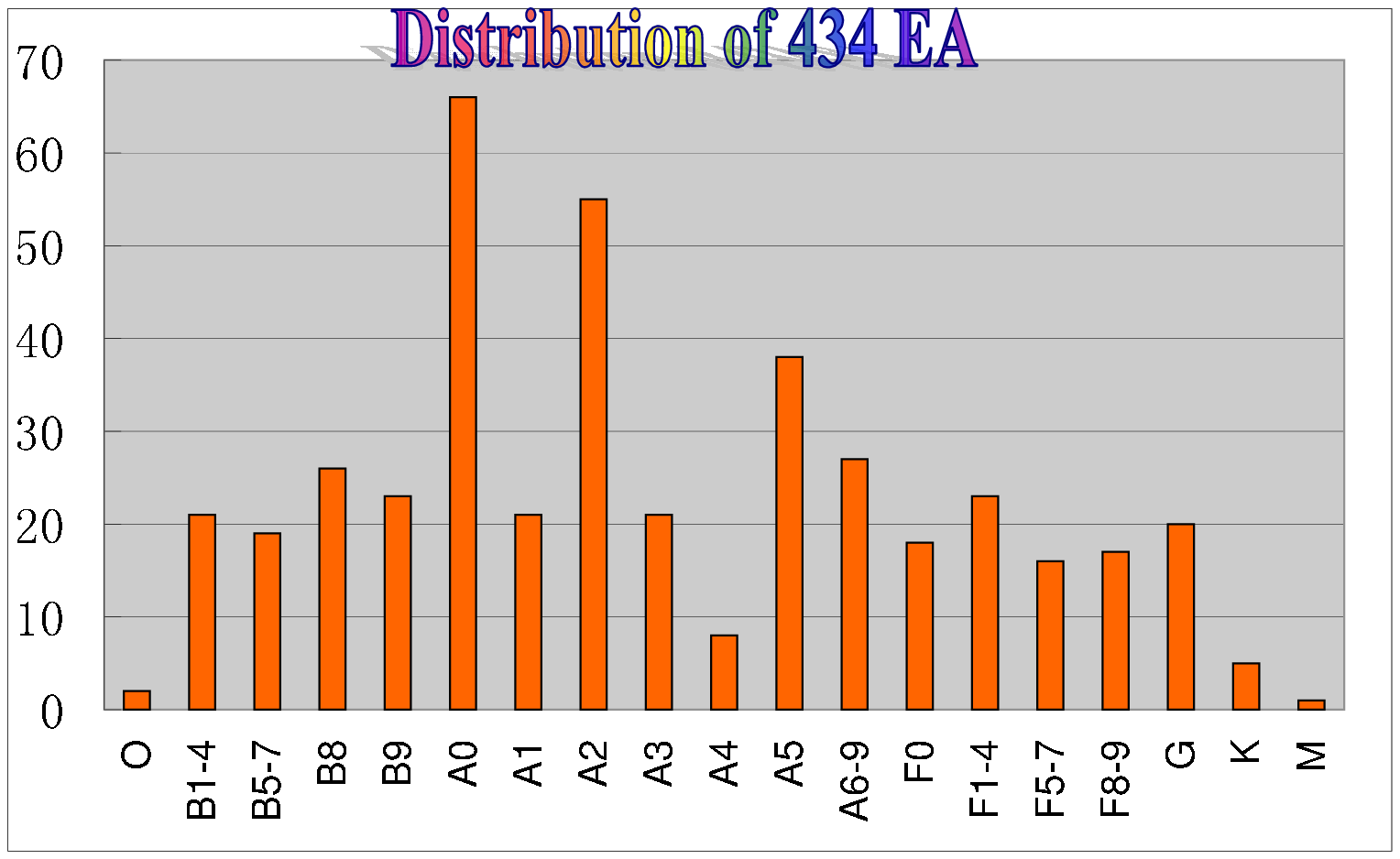}
   \caption{Distribution of the identified 434 Algol-type eclipsing binaries over spectral types.    }
   \label{Fig:EA-dist}
\end{figure}

\subsubsection{$\gamma$ Dor Pulsators in Binaries}
$\gamma$ Doradus-type stars (GDOR, prototype $\gamma$ Dor) are early type F dwarfs showing multiple periods
       from several tenths of a day to slightly in excess of one day.
       Amplitudes usually do not exceed 0.1 mag.  Presumably low degree gravity-mode
       non-radial pulsators and high-radial-order $g$-mode pulsators.
       Presently feasible driving mechanism is flux-blocking mechanism at the base of
       their relatively thin convective envelopes.
They usually have pulsation constants higher than 0.23\,d.
$\gamma$ Dor and $\delta$ Sct stars have commensurate pulsational periods.

\subsubsection{sdB/sdO Pulsators in Binaries}
Pulsating subdwarf B (sdB) variable stars
are low-mass ($\sim$0.5 M$_{\sun}$) core helium-burning horizonal branch stars
  with very thin outer hydrogen layers, making them quite luminous. They are evolved,
  compact (typical values $\log g \sim 5.8$) and hotter ($T_{\rm eff}\ga$ 20\,000\,K) B subdwarfs.
Since its discovery in 1997, over 30 of the sdBV stars
have been identified to be multimode pulsators, with typical pulsation periods of 100--250
seconds in a total range of about 60--600 seconds, and with pulsation amplitudes
generally less than a few hundredths of a magnitude. These pulsating sdB stars
are officially V361 Hya stars, which were commonly known as EC 14026 stars
after the prototype and referred to as sdBV stars.

\subsubsection{RR Lyr Pulsators in Binaries}
RR Lyrae stars (RR) are variables of the RR Lyrae type,
they are old population II, mostly found in globular clusters,
       which are radially-pulsating giant stars with spectral types in A7--F5,
       having amplitudes $\Delta V \sim$ 0\fm3--2\fm0 and periods in 1.5--24\,hr.
       Cases of variable
       light-curve shapes as well as variable periods are known. If
       these changes are periodic, they are called the ``Blazhko effect."
       The majority of these stars belong
       to the spherical component of the Galaxy; they are present, sometimes in
       large numbers, in some globular clusters, where they are known as
       pulsating horizontal-branch stars. Like Cepheids, maximum
       expansion velocities of surface layers for these stars practically
       coincide with maximum light. They are further classified into four subgroups:
(1) RR(B):~ RR Lyrae variables showing two simultaneously operating pulsation
       modes, the fundamental tone with the period $P_0$ and the first
       overtone, $P_1$ (AQ Leo). The ratio $P_1/P_0$ is approximately 0.745;
(2) RRab:~  RR Lyrae variables with asymmetric light curves (steep ascending
       branches), periods from 0.3 to 1.2 days, and amplitudes from 0.5
       to 2 mag in $V$; pulsating in fundamental mode.
(3) RRc:~  RR Lyrae variables with nearly symmetric, sometimes sinusoidal, light
       curves, periods from 0.2 to 0.5 days, and amplitudes not greater
       than 0.8 mag in $V$ (e.g. SX UMa). Overtone pulsators.
(4) RRd:~ RR Lyrae pulsators in first overtone and fundamental double radial modes.

\subsubsection{SX Phe-type Pulsators in Binaries}
Phenomenologically, these SXPHE resemble DSCT variables and
       are pulsating subdwarfs of the spherical component, or old disk
       galactic population, with spectral types in the range A2--F5. They
       may show several simultaneous periods of oscillation, generally in
       the range 0.04--0.08 days, with variable-amplitude light changes
       that may reach 0.7\,mag in $V$ band.
       These stars are Pop.\,II, metal-poor with high spatial motions,
       mostly in blue-straggler region in globular clusters.
       Multiperiodicity and nonradial pulsational contents are discovered recently
       in some of them.

\subsubsection{WDA/WDB Pulsators in Binaries}
White dwarf (WD) pulsators (showing absorption lines with FWHM $>$1500 km\,s$^{-1}$)
  with Balmer lines only (WDA) or white-dwarf white-dwarf binaries (WDB).
Binaries consisting of sdB and WD are listed together under sdB type.

\subsubsection{Oscillating Red Giant (Branch) Stars in Eclipsing Binaries}
Solar-like oscillations have been identified in 15 more red-giant branch (RGB) stars
belonging to eclipsing binary systems in {\it Kepler} data (Gaulme et al. 2013,2014).
The first detection was the 408-day period system KIC 8410637 (Hekker et al. 2010).
So far, all the stars known to both display acoustic modes and belong to EBs are
red-giants. We group them as red giants in eclipsing binaries (hereafter RGEBs or RG/EBs),
which span a range of orbital eccentricities, periods, and spectral types F, G, K, and M
for the companion of the red giant.

\subsubsection{Other Pulsators in Binaries}
In this contribution, we also collected
a few other types of pulsating stars in EB systems,
including three Wolf-Rayet stars, one BY Dra-type star,
and those eclipsing cataclysmic variables (CVs),
especially the subgroup of post-common envelope binaries known or suspected to possess planets.
CVs are a class of interacting binary star system which
display a huge diversity of physical phenomena.
The majority of them are composed of a white dwarf and a low-mass and
unevolved secondary star, plus an accretion disc through which material passes
from the secondary star (the donor) to the WD.
The importance of eclipsing CVs lies in the information which can be extracted from them: detailed modelling of their eclipses allows one to obtain
the basic physical properties of the system, including the masses and radii of the stellar
components. Such information is valuable in understanding the evolution of CVs.

\section{Statistics and Open Questions}
\label{sect:discussion}
Based on the catalog, several statistics were made in Table\,1 and Figs.3--6.
We address such stellar systems' observational implication to stellar evolutionary theory
by gathering the interesting topics and open questions from publications (Lampens 2006 and others)
as followings:
\begin{enumerate}
  \item How can binarity modify the pulsation properties? in what manner?
  e.g. how binary tidal interactions affect pulsations when compared to the single-star case.

  \item How can binarity/multiplicity help to identify the pulsation modes?
  Regarding that the amplitude and phase of the pulsating mode change during an eclipse (Breger 2005),
  the eclipse mapping technique was attempted by e.g. Reed et al.\,(2005), B\'{\i}r\'{o} \& Nuspl (2011) and
  the direct fitting of spherical harmonics by Latkovi\'{c} \& B\'{\i}r\'{o}\,(2008).

  \item Can we understand stellar pulsations when other processes
  (e.g. mass exchange/loss) are also present?

  \item What is the link between orbital motion, rotation and pulsation?

  \item We need improved models which can take into account deformed stellar shapes,
  including rotational and tidal distortions.

  \item How to discriminate properly between forced oscillations and modified
  or unaffected free oscillations in close binaries?

  \item Possible small cyclic variations of the oscillation frequencies,
  due to variable shape of the star in close eccentric-orbit binaries, and the
  light-time effect in the wide ones.

  \item binary constraints for asteroseismology of pulsating stars,
  e.g. the studies by Creevey (2008) and Creevey et al. (2011); establishing pulsating binary models,
  e.g. the work by Nie et al.(2010).

  \item search for solar-like $p$-mode oscillations in eclipsing binary systems.

  \item search for solar-like oscillations in metal-poor stars.

  \item search for pulsating M, K giants and subgiant stars.

  \item search for extra-solar planets orbiting a pulsator: eclipsing binaries
  consisting of planetary companions have been found from high-precision photometry,
 e.g. Silvotti et al.(2007), Qian et al.(2010).

 \item search for extragalactic eclipsing binaries containing pulsating
 stars (e.g. RR Lyr and Cepheids).
\end{enumerate}

Recalling the open questions, progress and prospects connecting oscillation and binarity,
the study of pulsating components in binaries becomes increasingly important.

\section{The Catalog}
\label{sect:catalog}

The catalog is arranged in different types of pulsating stars.
Entries for member stars in each group are listed in the order of ascending Right Ascension.
The orbital and pulsational data were adopted from literature.
FK5 coordinates (equinox=2000.0), spectral types and \emph{B,V,J,H,g,r} magnitudes when unavailable in
publications were adopted from the SIMBAD astronomical database\footnote{http://simbad.u-strasbg.fr/simbad/}
and other databases on the Internet.

\pagestyle{headings}
\paperheight 297 mm
\paperwidth 210 mm
\textheight 25.18 cm
\textwidth 16.90 true cm
\headheight 10pt
\headsep 10pt
\footskip 20pt
\marginparsep 10pt
\marginparwidth 20pt
\voffset = 0.05truecm
\hoffset = -0.5truecm

\baselineskip=4.2mm   
\setlength\topmargin{-8mm}  

\markboth{Zhou A.-Y.}{Catalog of Oscillating Binaries}

\begin{acknowledgements}
I appreciate Dr. Laszlo Szabados' comments, which urges me to finish cataloging
as complete as possible.
Thanks to Dr. Yang Y.-G. for providing the LaTeX format table of the 435 Algol-type eclipsing binaries.
This research was funded by the National Natural Science Foundation of China (NSFC).
The project is co-sponsored by the Scientific Research Foundation
for the Returned Overseas Chinese Scholars, State Education Ministry.
This research has made use of the SIMBAD database, operated at CDS, Strasbourg, France.
\end{acknowledgements}

\bigskip

\appendix

\section{Catalog of Identified Algol-type Eclipsing Binaries}

For the convenience of selecting a candidate EA to search for pulsation, currently confirmed
435 Algol-type eclipsing binaries are cataloged. As seen in Fig.\,\ref{Fig:HR-diagram},
pulsation could be excited in almost everywhere across the H--R diagram,
so spectral types is not an exclusive criterion in the selection of candidate pulsators.
Entries for member stars are listed in the order of ascending Right Ascension.
Some oEA systems have been included in this collection.
Other data and references to stars are not given for simplicity.
These materials are an update to the 434 entries given in the post-doctoral report of Dr. Yang Y.-G. (2010).

Besides this catalog, several catalogs from various survey projects are
available for reference. For instance,
the OGLE catalog of 2580+1351 EBs in LMC and SMC (Wyrzykowski et al. 2003, 2004),
11099 EBs in ASAS catalog (Paczynski et al.\,2006);
a catalog of 773 EBs in the TrES survey (Devor et al.\,2008;
\url{https://www.cfa.harvard.edu/~jdevor/Catalog.html});
{\it Kepler} EB stars catalog of 1879 EBs in its first data release (
\url{http://astro4.ast.villanova.edu/aprsa/kepler}; Prsa et al.\,2011) and
the second data release has increased the total number of identified EBs to 2165 (\url{http://keplerEBs.villanova.edu}; Slawson et al. 2011, Matijevic et al. 2012).
We investigated these catalogues and summarize them in Table~\ref{Tab:stat-EA}.
In addition, the fruitful searches (e.g. Mkrtichian et al.\,2002, Kim et al.\,2003) and
those fruitless attempts (e.g. Pazhouhesh et al. 2009)
can be referenced in selecting candidates.

\clearpage
\voffset = -0.25truecm
\hoffset = -0.05truecm

\begin{landscape}
\begin{table}     
\begin{center}
\caption[c]{~~A statistics on Algol-type eclipsing binaries. }
   \label{Tab:stat-EA}
   \vspace{1.68mm}
\begin{tabular}{crrll}
\noalign{\smallskip}\hline
       Catalog      & EA      & EB/EV     & Release Date                &References  \\
\hline\noalign{\smallskip}
(1)~ Catalog \#1  &2741       &47968 EV   & 2014.03, GCVS V4.2          & \href{http://www.sai.msu.su/gcvs/cgi-bin/search.htm}{\vp http://www.sai.msu.su/gcvs/cgi-bin/search.htm} \\
                  &4866       &47968 EV   & 2014.03, GCVS V4.2          & \href{http://www.sai.msu.su/gcvs/gcvs/iii/iii.dat}{\vp http://www.sai.msu.su/gcvs/gcvs/iii/iii.dat} \\
(2)~ Catalog \#2  &3395       &45835 EV   & 2012.03, HEASARC\_GCVS version& \href{http://heasarc.gsfc.nasa.gov/W3Browse/all/gcvs.html}{\vp http://heasarc.gsfc.nasa.gov/W3Browse/all/gcvs.html} \\
(3)~ Catalog \#3  &4062       & 7179 EV   & 2013.10,                    & Avvakumova et al. 2013, \href{http://dx.doi.org/10.1002/asna.201311942}{AN}, \href{http://adsabs.harvard.edu/abs/2013AN....334..860A}{\vp 334, 860}\\
(4)~ Catalog \#4  &3538       & 6330 EB   & 2005.09, from GCVS          & Malkov et al. 2006, \href{http://dx.doi.org/10.1051/0004-6361:20053137}{A\&A}, \href{http://adsabs.harvard.edu/abs/2006A\%26A...446..785M}{\vp 446, 785}\\
(5)~ Catalog \#5  & ---       & 2065 EB   & 2013.11, Kepler EB          & Kepler Eclipsing Binaries [Revision 3(beta)], \href{http://keplerebs.villanova.edu}{\vp http://keplerebs.villanova.edu}\\
(6)~ Catalog \#6  & ---       &26121 EB   & 2011 OGLE-III               & EB in LMC (Graczyk et al. 2011, Acta Astron., \href{http://adsabs.harvard.edu/abs/2011AcA....61..103G}{\vp 61, 103} \href{http://arxiv.org/abs/1108.0446}{arXiv:1108.0446})\\
\end{tabular}
   \end{center}
\vspace{-3mm}\hspace{8mm} \dag: The missing stars in the late catalogs were not investigated. EV=Eclipsing Variables.\\
\vspace{-0.0mm}\hspace{8mm}\ddag: Up-to-date version of this table at \url{http://www.chjaa.org/COB/}\\
\end{table}
\end{landscape}

\clearpage
\thispagestyle{plain}
\pagestyle{empty}
\textheight 25.18 cm
\textwidth 19.66 cm     
\hoffset -1.75 truecm   
\voffset -0.15 truecm   

\begin{deluxetable}{clclrrrrll}
\label{Tab:catalog-p1}
\tabletypesize{\footnotesize}  
\rotate
\tablenum{2}
\tablecolumns{10}
\tablewidth{0pt}    
\tablecaption{Catalog of Pulsators in Binary and Multiple Stellar Systems (web version: \url{http://www.chjaa.org/COB/})}
\tablehead{\colhead{Type of}                           &
           \multicolumn{1}{c}{Binary or}               &
           \colhead{RA~~~~~~~~~DEC}                    &
           \colhead{Sp.(A+B)}                          &
           \colhead{$\langle V \rangle$}               &
           \colhead{$\langle B \rangle$}               &
           \colhead{$P_{\rm orb}$}                     &
           \colhead{$P_{\rm pul}$}                     &
           \colhead{Comments}                                  &
           \colhead{References} \\

           \colhead{Pulsators}                                 &
           \colhead{Multiple}                                  &
           \colhead{hh:mm:ss~~~~~$^{\circ}:~^{\prime}:~^{\prime\prime}$}                                          &
           \colhead{}                                          &
           \colhead{(mag)}                                          &
           \colhead{(mag)}                                          &
           \colhead{(day)}                                     &
           \colhead{(day)}                                     &
           \colhead{}                                          &
           \colhead{}                                          }
\startdata
{\bf Cepheids:} & CG Cas  & 00:00:59~~~+60:57:32 & F5             & 11.37 &12.59 &  &        4.3652 & SB &\cite{272}\\
\cline{2-10}
{\bf Galactic}
                & SY Cas  & 00:15:10~~~+58:25:27 & F5             & 9.92  & 10.69 & &    4.0738& SB &\cite{275}\\
\cline{2-10}
{\bf DCEP}      & DL Cas  & 00:29:58~~~+60:12:43 & F7.5Ib + B9V   & 8.71  &  9.79  & 684.39 & 7.9983 &  SB+orbit, in NGC 129&\cite{16,65,99,118,196}\\
\cline{2-10}
                & BY Cas  & 01:47:12~~~+61:25:21 & F5             & 10.41 & 11.5   & 563 &  3.2211& SB+orbit, visual &\cite{104,105,106,280}\\
\cline{2-10}
                & CI Per  & 02:05:02~~~+57:08:34 & ---            & 12.7  & 13.00 & & 3.2961 & SB &\cite{221}\\
\cline{2-10}
                & UX Per  & 02:13:07~~~+58:04:48 & F5 +           & 11.0  & 11.3 & &  4.5709& SB,visual &\cite{277,280}\\
\cline{2-10}
               &V440 Per  & 02:23:52~~~+55:21:53 & F7Ib +         &  6.3  & 7.11& & 7.5683 & SB, unconfirmed,=ADS 1820 &\cite{104,107}\\
\cline{2-10}
                & VY Per  & 02:27:35~~~+58:55:02 & F5 +           & 11.26 & 12.79 & &   5.5335& SB &\cite{274}\\
\enddata

\tablenotetext{a}{Candidates are extracted from Simbad database without literature check.
There are no lines between candidates entries. }
\tablenotetext{b}{References numbers are accordingly linked to their entries in the list of references
at the end of the paper.}
\tablenotetext{c}{Check updates of this table on-line at: \url{http://www.chjaa.org/COB/}.
You are welcome to reporting any mistakes/typos as well as new members of
this category of oscillating binaries to the author at {\vp aiying@nao.cas.cn}.}

\end{deluxetable}

\clearpage
\voffset = -0.10truecm
\hoffset = 0.0truecm
\pagestyle{headings}
\markboth{Zhou A.-Y.}{Catalog of Oscillating Binaries (\url{http://www.chjaa.org/COB/})}

\input{EA435.tex}           

\clearpage
\voffset = -0.10truecm
\hoffset = -0.15truecm

\vspace{19.0cm}
{\it This manuscript was prepared in RAA format, but it is not yet accepted for publication.}

\label{lastpage}

\end{document}

%% file: EA435.tex
\begin{table}[thc]
\caption{~~~435 Identified Algol-type Eclipsing Binaries (web version: \url{http://www.chjaa.org/COB/}).}
\label{Tab:EA435}
\centering
\begin{tabular}{lllrcrcr}
\noalign{\smallskip}\hline
No.    & Stars  & Sp. Type   &RA (2000) & DEC(2000) & $\langle V \rangle$ &$Min.{\rm I}_{\textrm{HJD}}$&$P_{\rm orb}$ (d) \\
\hline
   1& DM Peg   & A8          &00:00:07.5 &$+$18:44:17.1 &10.80 &2425853.4700  &    2.5890 \\
   2& TW And   & F0V:        &00:03:18.2 &$+$32:50:45.1 & 9.08 &2443790.4490  &    4.1228 \\
   3& SX Cas   & G6          &00:10:42.1 &$+$54:53:29.4 & 8.96 &2417983.4300  &   36.5668 \\
   4& GT Cas   & A0          &00:13:29.4 &$+$58:16:58.0 &11.90 &2427341.4250  &    2.9898 \\
   5& TV Cas   & B9V         &00:19:18.7 &$+$59:08:20.6 & 7.22 &2443043.6265  &    1.8126 \\
   6& FS Cas   & F2          &00:24:44.2 &$+$57:18:37.0 &13.90 &2428038.5400  &    3.0809 \\
   7& V380 Cas & A0          &00:31:27.2 &$+$73:43:27.0 &10.40 &2425645.5050  &    1.3573 \\
\hline
\end{tabular}
\label{Tab A-11}
\end{table}

\addtocounter{table}{-1}
\begin{table}
\caption{~~ --- Continued}
\centering
\begin{tabular}{lllrcrcr}
\noalign{\smallskip}\hline
No.    & Stars  & Sp. Type   &RA (2000)  & DEC(2000) &$\langle V \rangle$ &$Min.{\rm I}_{\textrm{HJD}}$&$P_{\rm orb}$ (d) \\
\hline
  51& $\lambda$~Tau& B3V     &04:00:40.8 &$+$12:29:25.0 & 3.34 &2421506.8506  &    3.9529 \\
  52& RW Tau   & B8Ve        &04:03:54.3 &$+$28:07:34.0 & 8.00 &2445684.2040  &    2.7688 \\
  53& XZ Per   & G1          &04:09:27.8 &$+$46:33:57.0 &11.40 &2443507.4774  &    1.1516 \\
  54& RV Per   & A0          &04:10:38.0 &$+$34:15:54.9 &10.23 &2442046.9210  &    1.9735 \\
  55& BZ Eri   & F2          &04:12:12.6 &$-$06:01:17.6 & 9.00 &2425558.4450  &    0.6642 \\
  56& VX Hyi   & F4          &04:14:56.1 &$-$73:22:02.0 &11.90 &2425481.5000  &    3.5360 \\
  57& BN Tau   & $-$         &04:15:28.3 &$+$30:40:59.3 &10.17 &2445236.5840  &    4.2543 \\
\hline
\end{tabular}
\label{Tab A-12}
\end{table}

\addtocounter{table}{-1}
\begin{table}
\caption{~~ --- Continued}
\centering 
\begin{tabular}{lllrcrcr}
\noalign{\smallskip}\hline
No.    & Stars  & Sp. Type   &RA (2000) & DEC(2000) & $\langle V \rangle$ &$Min.{\rm I}_{\textrm{HJD}}$&$P_{\rm orb}$ (d) \\
\hline
 101& AK Aur   & A1          &06:39:17.3 &$+$31:36:46.0 &10.55 &2428126.4810  &    2.3814 \\
 102& AE Gem   & B9          &06:47:36.9 &$+$28:55:16.0 &10.50 &2435893.2810  &    1.4568 \\
 103& RX Gem   & A2          &06:50:11.6 &$+$33:14:21.0 & 9.21 &2440555.7820  &   12.2087 \\
 104& AF Gem   & A0          &06:50:39.7 &$+$21:21:56.0 &10.54 &2427162.3095  &    1.2435 \\
 105& RU Mon   & B9pv        &06:54:12.3 &$-$07:35:45.0 &10.33 &2441743.2900  &    3.5846 \\
 106& AU Mon   & B5          &06:54:54.7 &$-$01:22:32.9 & 8.17 &2442801.3752  &   11.1130 \\
 107& V745 Mon & B8          &06:55:11.9 &$-$09:19:16.2 & 7.69 &2448502.7700  &    3.9744 \\
\hline
\end{tabular}
\label{Tab A-13}
\end{table}

\addtocounter{table}{-1}
\begin{table}
\caption{~~ --- Continued}
\centering 
\begin{tabular}{lllrcrcr}
\noalign{\smallskip}\hline
No.    & Stars  & Sp. Type   &RA (2000) & DEC(2000) & $\langle V \rangle$ &$Min.{\rm I}_{\textrm{HJD}}$&$P_{\rm orb}$ (d) \\
\hline
 151& CW Vel   & B5III/IV    &09:02:21.3 &$-$52:50:28.9 & 9.04 &2444248.7584  &    2.3609 \\
 152& RX Hya   & A2          &09:05:41.2 &$-$08:15:39.2 & 8.90 &2443447.7000  &    2.2816 \\
 153& WW Cnc   & A8          &09:09:46.5 &$+$30:26:46.0 &10.50 &2427133.4590  &    1.1160 \\
 154& DN Vel   & A0III/IV    &09:19:37.7 &$-$45:40:47.8 & 9.38 &2428450.6900  &   12.8977 \\
 155& VZ Leo   & A5          &09:26:49.2 &$+$16:36:15.0 &11.80 &2431164.3160  &    1.0899 \\
 156& RX Vel   & A2          &09:27:32.6 &$-$51:04:15.5 & 9.68 &2429243.2720  &    3.1118 \\
 157& TY Hya   & A0          &09:29:02.8 &$+$05:34:32.0 &10.50 &2434478.4640  &    4.6611 \\
\hline
\end{tabular}
\label{Tab A-14}
\end{table}

\addtocounter{table}{-1}
\begin{table}
\caption{~~ --- Continued}
\centering 
\begin{tabular}{lllrcrcr}
\noalign{\smallskip}\hline
No.    & Stars  & Sp. Type   &RA (2000) & DEC(2000) & $\langle V \rangle$ &$Min.{\rm I}_{\textrm{HJD}}$&$P_{\rm orb}$ (d) \\
\hline
 201& BD Vir   & A8V         &13:26:40.6 &$-$16:06:16.4 & 9.96 &2442538.4130  &    2.5485 \\
 202& V380 Cen & B4V         &13:27:23.4 &$-$61:52:26.3 & 9.70 &2427807.6010  &    1.0872 \\
 203& V402 Cen & A0          &13:28:23.3 &$-$63:35:02.4 & 9.30 &2428330.3830  &    3.7198 \\
 204& RU UMi   & A2          &13:38:56.6 &$+$69:48:12.0 &10.00 &2441596.3365  &    0.5249 \\
 205& SY Cen   & A5          &13:41:51.6 &$-$61:46:10.1 &10.21 &2427892.6234  &    6.6314 \\
 206& T Cir    & B7II/III    &13:43:24.0 &$-$65:28:31.8 & 9.69 &2429095.5860  &    3.2984 \\
 207& SX Hya   & A5IV/V      &13:44:37.4 &$-$26:46:48.3 & 8.60 &2444344.4510  &    2.8957 \\
\hline
\end{tabular}
\label{Tab A-15}
\end{table}

\addtocounter{table}{-1}
\begin{table}
\caption{~~ --- Continued}
\centering 
\begin{tabular}{lllrcrcr}
\noalign{\smallskip}\hline
No.    & Stars  & Sp. Type   &RA (2000) & DEC(2000) & $\langle V \rangle$ &$Min.{\rm I}_{\textrm{HJD}}$&$P_{\rm orb}$ (d) \\
\hline
 251& V441 Oph & A0          &17:20:52.7 &$-$17:20:05.3 &11.60 &2435663.8210  &    3.0585 \\
 252& DW Aps   & B6III       &17:23:30.0 &$-$67:55:44.8 & 8.88 &2439209.5020  &    2.3129 \\
 253& V535 Oph & A3          &17:32:10.0 &$-$29:25:56.0 &11.30 &2427959.5050  &    6.0553 \\
 254& RV Oph   & A3V         &17:34:34.9 &$+$07:14:49.2 & 9.42 &2423997.3833  &    3.6871 \\
 255& RW Ara   & A1IV        &17:34:49.2 &$-$57:08:50.6 & 8.67 &2441861.8801  &    4.3672 \\
 256& SZ Her   & F0Vv        &17:39:36.8 &$+$32:56:46.6 & 9.95 &2441864.3052  &    0.8181 \\
 257& V806 Oph & F8          &17:40:56.0 &$-$16:57:35.0 &11.60 &2434216.6700  &   15.4065 \\
\hline
\end{tabular}
\label{Tab A-16}
\end{table}

\addtocounter{table}{-1}
\begin{table}
\caption{~~ --- Continued}
\centering  
\begin{tabular}{lllrcrcr}
\noalign{\smallskip}\hline
No.    & Stars  & Sp. Type   &RA (2000) & DEC(2000) & $\langle V \rangle$ &$Min.{\rm I}_{\textrm{HJD}}$&$P_{\rm orb}$ (d) \\
\hline
 301& BC Her   & A2          &18:50:41.2 &$+$12:29:46.0 &11.80 &2436483.4240  &    3.0873 \\
 302& HS Her   & B6III       &18:50:49.8 &$+$24:43:12.0 & 8.50 &2452856.3646  &    1.6374 \\
 303& BF Dra   & F8          &18:50:59.3 &$+$69:52:57.0 & 9.82 &2447276.3948  &   11.2111 \\
 304& BS Sct   & A7          &18:52:05.8 &$-$06:14:37.0 &11.00 &2440148.6200  &    3.8210 \\
 305& U Sct    & A9          &18:54:27.2 &$-$12:36:35.4 & 9.88 &2444468.6658  &    0.9550 \\
 306& CT Sct   & B9          &18:54:21.6 &$-$06:00:16.8 &10.02 &2428727.3740  &    4.9536 \\
 307& KP Aql   & A3          &19:02:29.9 &$+$15:48:01.1 & 9.70 &2452500.5670  &    3.3675 \\
\hline
\end{tabular}
\label{Tab A-17}
\end{table}

\addtocounter{table}{-1}
\begin{table}
\caption{~~ --- Continued}
\centering 
\begin{tabular}{lllrcrcr}
\noalign{\smallskip}\hline
No.    & Stars  & Sp. Type   &RA (2000) & DEC(2000) & $\langle V \rangle$ &$Min.{\rm I}_{\textrm{HJD}}$&$P_{\rm orb}$ (d) \\
\hline
 351& QY Aql   & F0          &20:09:28.8 &$+$15:18:44.7 &11.40 &2448507.1600  &    7.2296 \\
 352& V346 Aql & A0V         &20:09:59.6 &$+$10:20:59.1 & 9.00 &2441918.3840  &    1.1064 \\
 353& V345 Cyg & A1          &20:10:10.5 &$+$30:28:55.0 &11.30 &2428635.5680  &    2.0755 \\
 354& UZ Sge   & A0          &20:12:16.0 &$+$19:20:56.0 &11.40 &2445861.4140  &    2.2157 \\
 355& KU Cyg   & F0Iab:      &20:12:45.1 &$+$47:23:41.0 &10.73 &2433884.8400  &   38.4393 \\
 356& V695 Cyg & K2II        &20:13:37.9 &$+$46:44:29.0 & 3.73 &2441470.0000  & 3784.3000 \\
 357& VW Cyg   & A3          &20:15:12.3 &$+$34:30:48.0 &10.25 &2441116.8678  &    8.4303 \\
\hline
\end{tabular}
\label{Tab A-18}
\end{table}

\addtocounter{table}{-1}
\begin{table}
\caption{~~ --- Continued}
\centering  
\begin{tabular}{lllrcrcr}
\noalign{\smallskip}\hline
No.    & Stars  & Sp. Type   &RA (2000) & DEC(2000) & $\langle V \rangle$ &$Min.{\rm I}_{\textrm{HJD}}$&$P_{\rm orb}$ (d) \\
\hline
 401& RT Lac   & G5          &22:01:30.7 &$+$43:53:26.0 & 9.00 &2444873.3648  &    5.0739 \\
 402& DO Peg   & B8          &22:07:31.3 &$+$06:09:23.0 &10.60 &2445211.5100  &    2.6139 \\
 403& CY Lac   & B5          &22:13:08.3 &$+$54:33:35.0 &11.50 &2428746.2190  &    2.7869 \\
 404& AT Peg   & A3/5V       &22:13:23.5 &$+$08:25:30.8 & 8.97 &2445219.8562  &    1.1461 \\
 405& AU Lac   & A5:         &22:15:09.6 &$+$48:42:55.0 &12.00 &2434195.4410  &    1.3924 \\
 406& EK Cep   & A1V         &22:18:45.6 &$+$56:07:33.9 & 7.88 &2439002.7240  &    4.4278 \\
 407& UW Lac   & A5:         &22:20:39.8 &$+$42:24:29.0 &11.40 &2437188.3350  &    5.2902 \\
\hline
\end{tabular}
\end{table}